\documentclass[a4paper,twoside]{article}

\usepackage{epsfig}
\usepackage{subcaption}
\usepackage{calc}
\usepackage{amssymb}
\usepackage{amstext}
\usepackage{amsmath}
\usepackage{amsthm}
\usepackage{multicol}
\usepackage{pslatex}
\usepackage{apalike}
\usepackage{algorithm2e}
\usepackage[bottom]{footmisc}
\usepackage{listings}
\usepackage{color}
\usepackage{SCITEPRESS}     

\definecolor{lightgray}{rgb}{.9,.9,.9}
\definecolor{darkgray}{rgb}{.4,.4,.4}
\definecolor{purple}{rgb}{0.65, 0.12, 0.82}

\lstdefinelanguage{JavaScript}{
  keywords={typeof, new, true, false, catch, function, return, null, catch, switch, var, if, in, while, do, else, case, break},
  keywordstyle=\color{blue}\bfseries,
  ndkeywords={class, export, boolean, throw, implements, import, this},
  ndkeywordstyle=\color{darkgray}\bfseries,
  identifierstyle=\color{black},
  sensitive=false,
  comment=[l]{//},
  morecomment=[s]{/*}{*/},
  commentstyle=\color{purple}\ttfamily,
  stringstyle=\color{red}\ttfamily,
  morestring=[b]',
  morestring=[b]"
}

\begin{document}

\title{Skeet: Towards a Lightweight Serverless Framework Supporting Modern AI-Driven App Development}

\author{\authorname{Kawasaki Fumitake\sup{1}, Shota Kishi\sup{1} and James Neve\sup{1}}
\affiliation{\sup{1}ELSOUL LABO B.V., Netherlands, Amsterdam}
\email{\{f.kawasaki, s.kishi\}@elsoul.nl, jamesoneve@gmail.com}
}


\abstract{The field of web and mobile software frameworks is relatively mature, with a large variety of tools in different languages that facilitate traditional app development where data in a relational database is displayed and modified. Our position is that many current frameworks became popular during single server deployment of MVC architecture apps, and do not facilitate modern aspects of app development such as cloud computing and the incorporation of emerging technologies such as AI. We present a novel framework which accomplishes these purposes, \textit{Skeet}, which was recently released to general use, alongside an initial evaluation. \textit{Skeet} provides an app structure that reflects current trends in architecture, and tool suites that allow developers with minimal knowledge of AI internals to easily incorporate such technologies into their apps and deploy them.}

\onecolumn \maketitle \normalsize \setcounter{footnote}{0} \vfill

\section{\uppercase{Introduction}}
\label{sec:introduction}

The use of artificial intelligence (AI)-based technologies has rapidly expanded in recent years \cite{makridakis17}. Where AI was previously a useful but relatively niche technology used for extracting information from large datasets, it has recently been more broadly incorporated into a variety of existing technologies, from self-driving cars \cite{badue21} to healthcare \cite{hamet17}. The technology is also easier for non-specialist software developers to access, with companies such as Google offering pre-trained models via \textit{Google Cloud Platform} (GCP) which allow the incorporation of AI-driven technologies without requiring specific knowledge of model internals or training methodologies.

In particular, the incorporation of AI systems into standard web app development is becoming increasingly universal. For example, many companies incorporate chatbot technologies pioneered by services such as \textit{ChatGPT}\footnote{https://chat.openai.com/} into their websites in order to assist users interactively \cite{adamopoulou20}. Recent rapid developments in \textit{Large Langauge Model} (LLM) technology have allowed AI-driven chatbots to communicate more naturally and become useful components of app development \cite{brown20}. Methods of communication involving customer service representatives are typically frustrating for users \cite{parasuraman91}, either often requiring long waiting periods on the phone or for an email response. Chatbots trained in specific domain knowledge can help reduce the volume of communication with human customer service representatives by providing more specific advice than an FAQ page in a format tailored to the user's own knowledge, communication style and language \cite{nicolescu22}.

Traditional software frameworks are not usually equipped to facilitate development of AI-driven web apps. Since the advent and popularity of \textit{Ruby on Rails}\footnote{https://rubyonrails.org/}, many frameworks in different languages have followed its example in adopting \textit{Model-View-Controller} (MVC) paradigm of software development as the base of the framework \cite{laaziri19}, with the assumption that models are classes that form a 1-to-1 relationship with relational database tables; controllers orchestrate data retrieval of objects derived from one or more models; and views display this data to users. This paradigm has been an effective method of software development for many years, being both flexible enough to support a wide variety of software apps, and rigid enough that it is easy for developers to quickly understand, edit and test code. Unfortunately, this paradigm is often not well suited to developing AI-driven web apps. These usually depend on faster read/write operations from data sources such as \textit{NoSQL} databases \cite{han11} or even data warehouses such as \textit{BigQuery}\footnote{https://cloud.google.com/bigquery}, which are not always easy to conform to the MVC convention of exact correspondence between a model and a relational database table. AI-driven apps also often depend on cloud infrastructure, communicating with hosted machine learning resources. Third party libraries can alleviate some of these frustrations, but after a certain point considering alternative paradigms becomes appropriate.

In this paper, we present a novel lightweight \textit{Typescript}\footnote{https://www.typescriptlang.org/} framework called \textit{Skeet}\footnote{https://github.com/elsoul/skeet-cli}, which solves these problems. \textit{Skeet} is based on development according to \textit{Domain-Driven Design} \cite{evans04} philosophies: it encourages developers to split their code into domains which are independent in terms of business logic, and provides them with facilities for cloud deployment with support for multiple datasources, including relational and NoSQL database reads and writes. It also incorporates a command-line interface (CLI) suite which helps developers include popular pre-trained models such as those from \textit{Vertex AI}\footnote{https://cloud.google.com/vertex-ai} in their apps without requiring specific domain knowledge. As of January $2024$, the framework is increasing in popularity, with $562$ \textit{Stars} on \textit{Github} and having been used in a number of small projects.

The novel contributions made by this paper are as follows:
\begin{enumerate}
    \item We present design goals for a modern framework that represent the needs of people whose development goals are misaligned with current popular frameworks, which natively facilitates the incorporation of AI and cloud services.
    \item We present the architecture of a novel app framework, \textit{Skeet}, including user-friendly documentation, which aims to solve these design goals.
    \item An initial implementation of this software framework is evaluated using survey data collected from users who are experienced in traditional software frameworks, and have attempted app development with \textit{Skeet}.
\end{enumerate}

\section{\uppercase{Vision and Design Goals}}
\label{sec:visiondesigngoals}

This section provides an overview of the original motivations and design goals for \textit{Skeet}. It contains a review of the state of the art of software frameworks including definition of terminology, the vision for a modern software framework \textit{Skeet}, and a set of concrete design principles that guided \textit{Skeet}'s development.

\subsection{Web Frameworks}

This section describes web frameworks conceptually and provides definitions of terminology used throughout this paper.

Web frameworks are suites of tools that facilitate the development of a web app. Frameworks generally anticipate that an app will use a particular development paradigm (for example \textit{MVC}) and provide specific tools and commands around this. For example, these tools often include:

\begin{itemize}
    \item Database management systems, commonly \textit{migrations} which allow reversible step-by-step database construction in different environments.
    \item \textit{Object-relational Mapping} (ORM) libraries to facilitate communication with data stores using code.
    \item Automatic code and test generation via a \textit{Command Line Interface} (CLI) suite for common tasks.
    \item Routing libraries to connect specific subdomains to executable functions which return \textit{HTML} or formatted data such as \textit{JSON}.
\end{itemize}

Existing frameworks are in practice classified along a scale of \textit{lightweight} to \textit{heavyweight}. Lightweight frameworks are minimalist, providing tools to create an app while allowing the developer to retain relative freedom of architecture design. An example of a lightweight framework in \textit{Ruby} is \textit{Sinatra}\footnote{https://github.com/sinatra/sinatra}, which provides minimal tools such as routing. A heavyweight framework often defines the structure of the app, and uses assumptions about the developer's use of this structure to provide more detailed tools and commands. For example, \textit{Ruby on Rails} extends its tool suite to security, boilerplate code and test generation and data extraction.

The framework examples above are used to solve the general problem of constructing a web app. In the absence of terminology, this paper defines them as \textit{generalist} frameworks. In contrast, \textit{specialist} frameworks attempt to solve a problem in a more specific domain. For example, e-commerce in the context of the Semantic Web \cite{li03} or medical image processing \cite{tournier19}.

\subsection{The Vision}
\label{sec:vision}

This section describes the original motivations for the development of \textit{Skeet}, which were threefold:

\begin{enumerate}
    \item Services such as \textit{ChatGPT} and \textit{Vertex AI} have been rising in popularity as a result of developments in machine learning algorithms and advances in cloud computing that have made training large scale models more accessible. Using these models as a base, construction of specific chatbots has become more accessible to a wide variety of developers. It would be particularly useful for individual developers to be able to rapidly construct chatbots from their company's data (for example, customer service communications data from an e-commerce app). A framework is needed which facilitates this through a simple CLI.
    \item There has recently been an expansion in popularity of using strongly typed languages which compile into \textit{Javascript} to write front-end and back-end code. For example, \textit{Scala.js}\footnote{https://www.scala-js.org/} and \textit{Dart}\footnote{https://dart.dev/} allow the compilation of their respective languages into \textit{Javascript}. This facilitates communication between front-end and back-end developers, which has traditionally been a pain point in organisations where the two departments may struggle with scheduling due to lack of understanding of the other's development processes. We consider that \textit{Typescript} has been popular for a long time as the base for front-end frameworks such as \textit{AngularJS}\footnote{https://angularjs.org/} and \textit{React}\footnote{https://react.dev/}, and is also suitable as a back-end language. \textit{Skeet} is intended to provide a tool suite for both back- and front-end developers to use \textit{Typescript} for their development, and thus reduce the intra-organisational communication burden.
    \item Skeet's serverless architecture should allow developers to quickly set up infrastructure without in-depth knowledge and support scalability. Traditional development approaches often require upfront design considerations for load balancing and resource allocation, which not only lead to higher design and fixed infrastructure costs but also limit the application's ability to scale efficiently. The serverless model is designed to automatically adjust to varying loads, enabling not just rapid but also scalable product development. This adaptability ensures that service owners pay according to resource usage, effectively solving longstanding issues related to infrastructure knowledge, management and cost.
\end{enumerate}

\subsection{Design Goals}
\label{sec:designgoals}

This section describes concrete design goals of \textit{Skeet}, which were distilled from the three overarching motivations described in section \ref{sec:vision}. As a framework, Skeet should:

\begin{enumerate}
    \item Allow full-stack or backend developers to create apps based on AI chatbots with minimal machine learning expertise. It should have a command line interface that allows for the implementation and fine tuning of various models with relatively little direct coding on the part of the developer themselves, allowing them to focus on preparing the data and building the back end and front end around the chatbot.
    \item Be a lightweight framework. While little machine learning expertise is assumed, we consider that a wide variety of apps can be built around chatbots, and therefore the framework should not place unnecessary restrictions on the developer. \textit{Skeet} does allow and indeed facilitates the use of a user's choice of front-end framework such as \textit{React} in addition, in order to shape design code.
    \item Support cloud infrastructure. Chatbots and other AI services are often processor- and memory-intensive technologies that require cloud computing for training and prediction. \textit{Skeet} provides a suite of tools for connecting to popular chatbot APIs such as \textit{ChatGPT} and \textit{Vertex AI}, fine tuning models and for deploying AI-driven apps directly on \textit{Google Cloud Platform}.
    \item Facilitate communication between back-end and front-end developers by reducing the burden of the language barrier. \textit{Skeet} assumes a back end written in \textit{Typescript}, which is by far the most popular front-end language thanks to frameworks such as \textit{React}, and therefore improves understanding of back-end technologies by front-end developers including AI-related components.
    \item Leverage Serverless Architecture for Scalability and Cost-Efficiency. \textit{Skeet}'s serverless framework eliminates the need for upfront infrastructure design, thereby reducing both time and cost. This serverless model inherently supports scalability, automatically adjusting to varying loads without requiring manual intervention. This not only enables rapid and scalable product development but also allows for a pay-as-you-go approach to resource usage. The serverless architecture makes it feasible for small teams or even individual developers to build and scale applications without the complexities of traditional infrastructure management.
\end{enumerate}

\section{Framework Design}
\label{sec:framework}

This section introduces the architecture of the \textit{Skeet} framework. It describes how the concepts introduced in the previous sections are used in terms of concrete implementation.

\subsection{Cloud Infrastructure Architecture}

\begin{figure*}[!htb]
    \center{\includegraphics[width=0.7\textwidth]{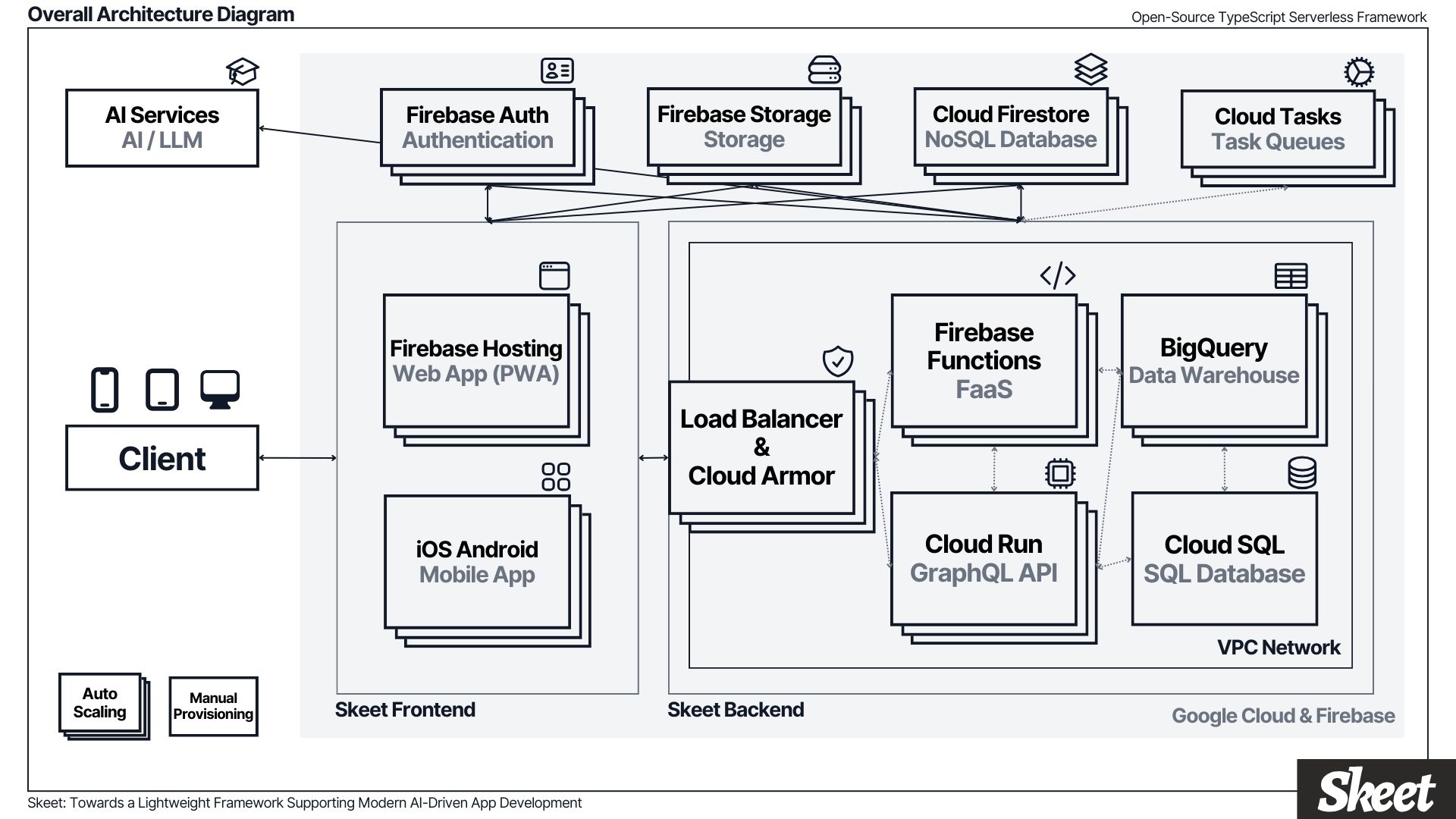}}
    \caption{An overview of the \textit{Skeet} infrastructure architecture.}
    \label{fig:infra-architecture}
\end{figure*}

Figure \ref{fig:infra-architecture} shows the overall structure of the \textit{Skeet} infrastructure architecture. As outlined in Section \ref{sec:introduction}, the framework is designed around the assumption that \textit{GCP} will be used for infrastructure, which allows communication between the backend and infrastructure to be handled by the framework code and CLI commands, simplifying the process for users and especially developers who are specialised in coding but less so in infrastructure design.

The architecture assumes a general use app which requires a front end, back end and data storage. The overall architecture is designed with the suggested best practices of \textit{Statelessness} and \textit{Immutability}\footnote{https://cloud.google.com/architecture/best-practices-for-operating-containers\#immutability}. 

The architecture incorporates appropriate data storage, with structured data in a \textit{Cloud SQL} instance, unstructured data accessible directly to the front end stored in a \textit{NoSQL} database and \textit{BigQuery} used as a data warehouse for analysis purposes. These instances are automatically set up as part of the \textit{Skeet} initialization process, but these steps can be omitted if the user desires to design their own cloud architecture.

Configuration parameters for database settings are defined by internal variables which are 
stored as \textit{Fireabse Secrets} when they are initiated. Users have the option to modify these configuration parameters manually, giving them fine-grained control over their infrastructure.

The workflow of a typical \textit{Skeet} app as shown in Figure \ref{fig:infra-architecture} is as follows. The client communicates with a frontend which is either a mobile app or a website. The main frameworks recommended by \textit{Skeet} and supported at the CLI generator level are \textit{React}\footnote{https://react.dev/} and \textit{React Native}\footnote{https://reactnative.dev/}, but other frontends can be used if the user has particular preferences. The frontend communicates directly with \textit{Firestore} in order to rapidly access data authorized for the user based on \textit{Firebase} login protocols. Where business logic or relational data is required, the front end accesses the back end via a load balancer. Back-end logic is expected to be hosted as a series of independent functions on \textit{Cloud Functions}

AI services are accessed via business logic stored on \textit{Cloud Functions}. Popular LLM-based chatbots such as \textit{OpenAI} and \textit{Vertex AI} can be accessed by registering API keys with the framework. The framework code automates API access and provides CLI code generators to quickly output the \textit{Typescript} code required.

\subsection{Back-End Architecture}

Back-end code in a \textit{Skeet} app is divided into \textit{Functions}. As a general rule, the flow of execution of back end code is as follows:

\begin{enumerate}
    \item The frontend calls a \textit{Skeet} function as an HTTP request.
    \item The function retrieves data from data sources, potentially including relational and \textit{NoSQL} data and responses from AI resources.
    \item The function performs business logic operations on the data, re-formats it for displayability and returns it as a JSON response.
\end{enumerate}

A Skeet app can have any number of \textbf{functions}, and users are encouraged to design fine-grained functions which separate business logic, following \textit{SOLID} design principles \cite{martin00} and in particular the \textit{Single Responsibility} principle, which states that each class should solve only one problem; broadly speaking the same can be said of a \textit{Skeet} function within the broader context of domain logic.

A new function can be generated by a CLI command \lstinline{skeet add functions --name} with initial configuration files and an \lstinline{src} directory for user-written code. A function is initialised with the following directories:

\begin{itemize}
    \item \textbf{lib} directory contains classes that perform business logic operations on data.
    \item \textbf{routings} directory provides HTTP endpoints which can be called by the front end to retrieve or mutate data.
    \item \textbf{scripts} directory contains code which executes scheduled operations independent of the HTTP endpoints. 
    \item \textbf{types} directory contains data storage classes related to front-end logic. Data from the database is reorganised from models into types before being returned.
\end{itemize}

Users can create blank functions for their own code, or they can create pre-build functions with scaffolding for the purposes of using AI tools such as chatbots. In the future, community users will be able to design pre-built functions that anyone can add to their project.

In addition to domain-specific code which is cleanly separated into functions, \textit{Skeet} also provides a \textbf{common} directory for cross-function code. This is deployed independently, can be accessed from all functions, and includes the following:

\begin{itemize}
    \item \textbf{models} directory provides data storage classes that correspond to units of structured data, such as relational database tables. It is likely that independent pieces of domain logic will require data from the same sources.
    \item \textbf{utils} directory contains code that executes common operations across business logic. This is intended for small, repeated functions that are independent of domain logic.
    \item \textbf{enums} directory contains domain-independent variable definitions.
\end{itemize}

The \textit{Skeet} CLI expects that users will organise their apps using this broad structure, and provides commands to generate scaffolding and tests under this assumption. However, as a lightweight framework, it does not enforce this structure as such; users are free to decide their own directory structure, and the framework will still function with the exception of the automatic code generation functions from the CLI.

\subsection{Skeet AI}

A key feature of \textit{Skeet} is the tools that facilitate incorporating AI and in particular chatbot features into web and mobile apps. Machine learning models are often difficult for developers to understand from first principles, and training and deploying original models requires considerable specialist knowledge in addition to, depending on the model and data, significant expenditure on computing resources.

\textit{Skeet} provides internal libraries and code scaffolding CLI commands for connecting to common AI services such as \textit{ChatGPT} and \textit{Vertex AI} such that users only have to input their API credentials in order to connect to and employ models in their apps.

In addition to back-end code scaffolding and endpoint generation, \textit{Skeet} also provides front-end resources to help users specifically with chatbot setup. Chatbot implementation can be time-consuming to implement for front-end developers. \textit{Skeet} provides front-end commands to assist with this process, so users can set up pre-trained or fine tuned implementations with minimal manual code entry.

Finally, \textit{Skeet} uses pretrained models to assist in automatic code generation. Employing the \textit{Prisma}\footnote{https://www.prisma.io/} ORM, it allows users to define their database tables using natural language, and converts them to migrations and model files automatically.

\section{\uppercase{Evaluation}}
\label{sec:evaluation}

This section describes the evaluation of an initial implementation of \textit{Skeet} based on the design goals outlined in Section \ref{sec:designgoals}.

\subsection{Survey Design}

An initial evaluation for the framework was conducted based on survey data collected to identify its successes and fix potential flaws. The framework is still in development, and this survey therefore represents a snapshot of early adopters, and is not intended as a rigorous and unbiased evaluation, which will be conducted when development is completed. The objectives of the survey were threefold:

\begin{enumerate}
    \item To establish the extent to which the initial release of \textit{Skeet} achieves the design goals outlined in Section \ref{sec:designgoals}.
    \item To find out what kind of developers the framework currently appeals to.
    \item To identify current issues and pain points for developers using the framework so they can be fixed.
\end{enumerate}

Survey questions were designed with these goals in mind. The questions can be split into two categories: questions to gather information about the participant, and questions to evaluate the framework. Where possible, questions were asked with answers on a \textit{Likert Scale} \cite{nemoto14} from $1$-$5$, which provides a quantitative measurement of the results, with two positive responses, a neutral response and two negative responses.

Information gathering questions were asked with the objective of tailoring future development and better understanding the participants' evaluative questions. These are listed below, with possible answers in brackets where answers were not free choice:

\begin{enumerate}
    \item Did you complete the tutorial and create a sample project using Skeet? (Yes / No)
    \item Which programming languages do you have experience with? (Checkbox Selection)
    \item Which web frameworks do you have experience with? (Checkbox Selection)
    \item How experienced are you with AI technologies? (Professional experience / Quite experienced / Some experience / Inexperienced / No experience)
    \item How experienced are you with setting up cloud architectures? (Professional experience / Quite experienced / Some experience / Inexperienced / No experience)
\end{enumerate}

Evaluation questions were asked with the objective of finding out participants' satisfaction with the framework as compared to other frameworks, and identifying specific pain points that they may have experienced. The questions with available answers in brackets are listed below.

\begin{enumerate}
    \setcounter{enumi}{5}
    \item How easy was it to create and run a new project in Skeet compared to other frameworks? (Very easy / Quite easy / Average / Quite difficult / Very difficult)
    \item Compared to other frameworks you’ve used in the past, how easy was it to create a chatbot using Skeet? (Very easy / Quite easy / Average / Quite difficult / Very difficult)
    \item Compared to other frameworks, how easy was it to understand the code architecture of Skeet? (Very easy / Quite easy / Average / Quite difficult / Very difficult)
    \item How easy was it to set up GCP services using Skeet’s CLI compared to the time taken to do it manually? (Very easy / Quite easy / Average / Quite difficult / Very difficult)
    \item How likely are you to consider using Skeet for a future project? (Very likely / Quite likely / Possible / Unlikely / Very unlikely)
\end{enumerate}

The survey also asked for qualitative comments from participants about the framework, with the following question: "Please write any comments you have about your experience using \textit{Skeet} below". The survey was conducted on \textit{Google Forms}, and collected email addresses but no other identifying information from participants. All quantitative responses were mandatory, and qualitative responses were optional.

\subsection{Results}

$136$ participants responded to the online survey, advertised on \textit{X} (formerly \textit{Twitter}), on the \textit{Discord} for the framework's development and on the website for the framework. Percentages in the text of this section are rounded to the nearest whole number.

The development experience of users was diverse, but answers indicated that most respondents came from a web development background. Most commonly (question $2$), users had experience developing in \textit{Python} ($62\%$), \textit{Javascript} ($58\%$) and \textit{Typescript} ($47\%$). Framework experience (question $3$) was similar, with \textit{ReactJS} ($51\%$), \textit{Ruby on Rails} ($41\%$) and \textit{Angular} ($37\%$) being significantly higher than other choices.

\begin{figure}[!htb]
    \center{\includegraphics[width=0.4\textwidth]{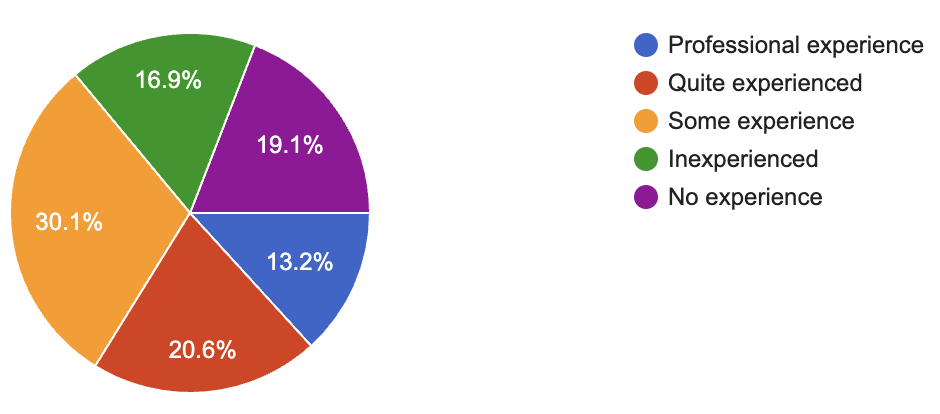}}
    \caption{Question 4 - How experienced are you with AI technologies?}
    \label{fig:results-ai}
\end{figure}

Questions $4$ and $5$ asked about users' previous experience with technologies outside of strictly development. Experience with AI development was of particular note, shown in Figure \ref{fig:results-ai}: $64\%$ of users expressed at least some experience, with $13\%$ having professional experience, which is larger than we anticipated given that the original objective of \textit{Skeet} is to assist developers who have very little experience with AI technologies. Experience with cloud architectures showed a similar distribution, with a slightly higher level of general experience: $68\%$ with at least some experience or more, $17\%$ with professional experience. This is not unsurprising, as cloud deployment is relatively common in the web development space. Both results, however, are slightly higher than originally anticipated, and possibly reflect the fact that early adopters of a framework are more likely to be experienced developers. Further user interviews focused specifically on an inexperienced audience might be illuminating.

$92\%$ of users who submitted the survey were able to complete the tutorial (question $1$). This indicates a high level of reliability across a wide variety of different development environments. One user who was not able to complete the tutorial submitted the error they encountered as a comment, which implied that mistakenly skipping steps in the tutorial was a possible reason users were unable to complete it. In order to address this, the steps that appear to be commonly skipped will be further emphasised in the tutorial, and error messages will be updated to indicate to users when steps might have been missed, and what should be done to correct this.

\begin{figure}[!htb]
    \center{\includegraphics[width=0.4\textwidth]{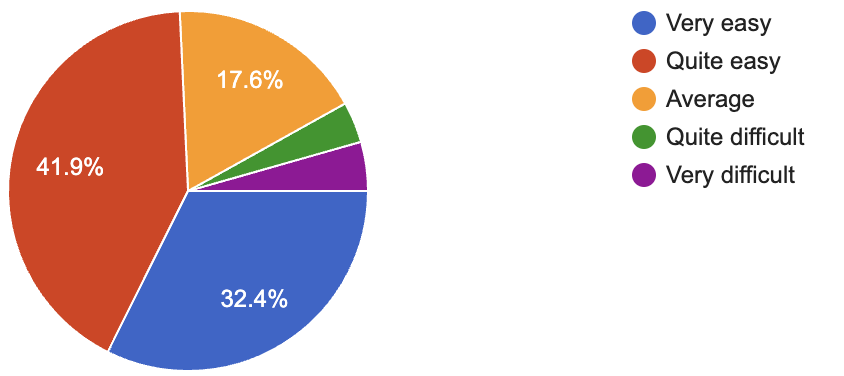}}
    \caption{Question 6 - How easy was it to create and run a new project in Skeet compared to other frameworks?}
    \label{fig:results-how-easy}
\end{figure}

$92\%$ of users who were able to complete the tutorial responded that being able to create and run a new project in \textit{Skeet} (question $6$) was average or easier than average, with $32\%$ answering that it was very easy (Figure \ref{fig:results-how-easy}). This is an encouraging distribution, as it indicates that a significant percentage of users who were inexperienced or have no experience with AI and cloud technologies found it quite easy or very easy, which was the original objective of framework development.

\begin{figure}[!htb]
    \center{\includegraphics[width=0.4\textwidth]{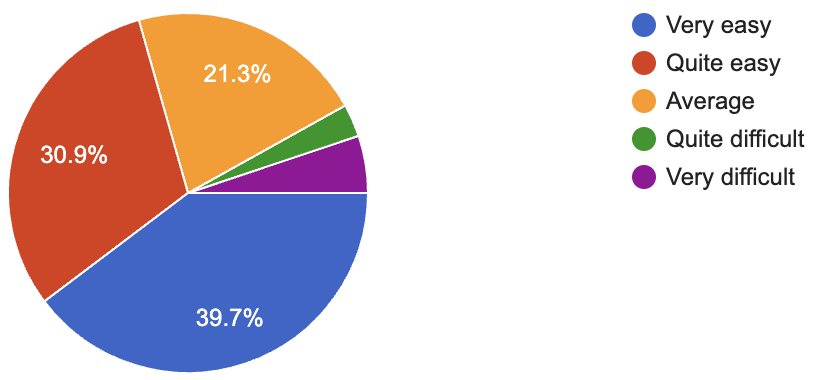}}
    \caption{Question 8 - Compared to other frameworks you’ve used in the past, how easy was it to create a chatbot using Skeet?}
    \label{fig:results-chatbot}
\end{figure}

Distributions for questions $8$ and $9$ were similarly skewed towards the positive end of the spectrum. $69\%$ of users indicated that it was either quite or very easy to create a chatbot using \textit{Skeet} compared to other frameworks (Figure \ref{fig:results-chatbot}) and $72\%$ said the same thing about their experience setting up cloud services in question $9$.

Qualitative results were similarly encouraging. $14$ comments were submitted, of which $12$ were positive but nonspecific (for example, ``The speed of development and support is fast, which is helpful for developers.''). One criticism submitted the error they encountered during the tutorial which, as discussed higher up in this section, came from skipping a tutorial step. The other criticism stated, ``It was difficult for beginners''. The implication here is unclear, but \textit{Skeet} is not designed for complete beginners to programming, so this user's experience may not be inconsistent with the design goals.

These results indicate that this early version of \textit{Skeet} is effectively solving the design goals outlined in Section \ref{sec:designgoals}. Users were consistently able to complete the tutorial, and those that did so indicated that their experience with the framework was positive, and they found it easier than other frameworks than they had used for the specific purposes of AI and cloud service integration. When development work is completed on the framework, more rigorous surveys will be conducted examining all aspects of the framework's ease of use.

\section{\uppercase{Conclusions}}

Our position is that many of the popular frameworks currently being used to develop apps were designed in a previous era, and methods of modern app development such as serverless architectures and the use of tools such as AI are not correctly served by these frameworks. Where they do provide these features, they are often added subsequently by third parties and therefore more difficult to use than those same features in a framework that had been designed with them in mind.

As a potential solution to this problem, this paper presented a lightweight framework, \textit{Skeet}. \textit{Skeet} is written in \textit{Typescript}, and allows immediate serverless deployment upon app creation, regardless of the level of infrastructure expertise of the user. The framework also allows simple incorporation of AI features such as chatbots based on pretrained models as CLI-generated functions. Section \ref{sec:designgoals} described specific design goals for the framework, and Section \ref{sec:framework} described how each section of the framework was developed and therefore achieved the outlined goals.

The framework was initially evaluated through collecting survey data from users of the framework, as described in Section \ref{sec:evaluation}. This evaluation indicated that this early version of the framework has been successful in achieving its design goals: a vast majority of the early adopters completed the tutorial, and found it easier than mainstream alternatives for accomplishing the objectives outlined in the initial design goals. 

For web development to continue to progress into the modern era, and for advanced techniques to continue to be accessible, we require new frameworks which allow non-specialist developers to incorporate these features while hiding some of the complexity involved in their implementation. \textit{Skeet} represents an innovative solution to this problem, and will continue to be developed with this goal in mind.

\section*{\uppercase{Acknowledgements}}

We would like to acknowledge and thank \textit{East Ventures}, \textit{MIRAISE} and \textit{Rikka} for supporting this research project.

\bibliographystyle{apalike}
{\small
\bibliography{enase_2024}}

\end{document}